# Spatial Correlation at the Boson Peak Frequency in Amorphous Materials

X. Y. Li[1]†‡, H. P. Zhang[2,3]†, S. Lan[4,5], D. L. Abernathy[6], C. H. Hu[1], L. R. Fan[1], M. Z. Li[2,7]*, X.-L. Wang[1,5,8,9]*

[1] Department of Physics, City University of Hong Kong; 83 Tat Chee Avenue, Hong Kong, China.

[2] Department of Physics, Beijing Key Laboratory of Opto-electronic Functional Materials and Micro-nano Devices, Renmin University of China; Beijing 100872, China.

[3] Songshan Lake Materials Laboratory; Dongguan 523808, China.

[4] Herbert Gleiter Institute of Nanoscience, School of Materials Science and Engineering, Nanjing University of Science and Technology; 200 Xiaolingwei Avenue, Nanjing 210094, China.

[5] Center for Neutron Scattering, City University of Hong Kong; 83 Tat Chee Avenue, Hong Kong, China.

[6] Neutron Scattering Division, Oak Ridge National Laboratory; Oak Ridge, Tennessee 37831, USA.

[7] Key Laboratory of Quantum State Construction and Manipulation (Ministry of Education), Renmin University of China; Beijing, 100872, China.

[8] Hong Kong Institute for Advanced Sciences, City University of Hong Kong; 83 Tat Chee Avenue, Hong Kong, China.

[9] City University of Hong Kong Shenzhen Research Institute; 8 Yuexing 1st Road, Shenzhen Hi-Tech Industrial Park, Shenzhen 518057, China.

† These authors contributed equally to this work.

‡ Present address: Department of Physics & Astronomy and Stewart Blusson Quantum Matter Institute, University of British Columbia; Vancouver V6T1Z4, Canada.

* Corresponding author. Email: maozhili@ruc.edu.cn, xlwang@cityu.edu.hk

**The Boson peak (BP), an excess of vibrational density of states, is ubiquitous for amorphous materials and is believed to hold the key to understanding the dynamics of glass and glass transition. Previous studies have established an energy scale for the BP, which is ~1-10 meV or ~THz in frequency. However, so far, little is known about the momentum dependence or spatial correlation of the BP. Here, we report the observation of the BP in model Zr-Cu-Al metallic glasses over a wide range of momentum transfer, using inelastic neutron scattering,**

**heat capacity, Raman scattering measurements, and molecular dynamics (MD) simulations. The BP energy is largely dispersionless; however, the BP intensity was found to scale with the static structure factor. Additional MD simulations with a generic Lennard-Jones potential confirmed the same. Based on these results, an analytical expression for the dynamic structure factor was formulated for the BP excitation. Further analysis of the simulated disordered structures suggests that the BP is related to local structure fluctuations (e.g., in shear strain). Our results offered insights into the nature of the BP and provide guidance for the development of theories of amorphous materials.**

Upon rapid cooling, liquids will solidify into non-equilibrium glassy or amorphous states[1–3]. As such, dynamic relaxation is a universal and intrinsic feature of glass and enables fluctuation and dissipation to occur[4,5]. Dynamic relaxation covers a broad range of time, length, and temperature scales, which in turn determines the properties and applications of glassy systems. As a result, there has been growing interest in the physics of glass[6–13]. Theoretically, the dynamical modes of glass can be clarified into two categories[4]. The first category pertains to transitions between adjacent basins or configurations within the potential energy landscape[14] encompassing the so-called α and β relaxation processes, while the second one involves the vibrational modes within a singular basin. The investigation of vibrational modes holds significant importance due to its connection with the transport properties such as thermal conductivity and heat capacity, which plays a crucial role in determining the performance of thermoelectric materials, as well as serving as a criterion for determining structural phase transitions in condensed matter[15]. In ordered crystals with long-range order, the vibrational modes manifest as plane-wave phonons, resulting in a vibrational density of states (VDOS) $g(\omega)$ that adheres to Debye-like behavior at low frequencies[16]. In contrast, the VDOS of amorphous materials generally departs from the Debye form, exhibiting an excess of states known as the BP[17]. Following decades of intensive investigation, a consensus has formed that the BP is linked to specific sorts of disorder[13,18–21]; however, the underlying structural mechanism of the BP has remained enigmatic and a subject of intense debate.

In history, the BP was first discovered in the 1950s in thermal conductivity and Raman spectra of vitreous $SiO_2$ by Berman[22] and Krishnan[23]. It was termed BP because the temperature dependence of the Raman intensity follows that of a harmonic oscillator characterized by the Bose factor. Since then, the BP has been widely observed in various amorphous materials, ranging from metallic glasses (MGs)[24,25], polymer and molecular glasses[18,26–29], oxide and chalcogenide glasses[17,20,30–32], ionic salt and bio-matter glasses[33,34], colloidal and granular glasses[21,35], model glasses[13,36–39], confined water[40,41], molecular crystal and strain glass[42,43], van der Waals liquids[44], and spin glass[45], in spite of the very different interactions and structures in these materials. This general and fascinating anomaly can be identified in the form of a broad peak in the specific heat data[46,47] as well as in the VDOS, which can be measured by Raman scattering[28,48], inelastic neutron scattering (INS)[18,49], and inelastic x-ray scattering (IXS)[26,50].

Considerable studies have been carried out to explore the physical origin of the BP[13,21,31,36–38,47,51–53]. In general, there are two schools of thought regarding the origin of the BP. In the first view, the BP is considered a fingerprint of glass, i.e., an excess mode arising from the disordered atomic packing in amorphous materials[13,36,51]. In the second view, the DOS of the amorphous system is regarded as a modification of the crystalline DOS due to a random fluctuation of the force constants[21,31,52]. The controversy surrounding the BP has lasted for decades, constituting one of the most fascinating and difficult puzzles in condensed matter physics and materials science.

In our view, this controversy is mainly caused by the incomplete definition of BP. In condensed matter, an elementary excitation is generally defined with two parameters, i.e., energy and momentum – their dynamic response functions describe the temporal and spatial correlation, respectively[16]. However, in literature, the BP was only defined with an energy value, where the spatial correlation was largely ignored. As a result, much of the efforts in the past have focused on the energy scale of the BP, while few studies paid attention to the momentum dependence or spatial correlation at the BP energy. The lack of knowledge about the spatial correlation of the BP has severely limited our understanding of the nature of the BP in amorphous materials. Here, we report a concerted investigation combining the powerful INS and MD simulations, to establish the spatial correlation of the BP in $Zr_{46}Cu_{46}Al_8$ and $Zr_{56}Cu_{36}Al_8$, two typical MGs with very different glass-forming abilities. The results are similar for both alloys and will be illustrated mainly with the data from Zr$_{46}$Cu$_{46}$Al$_8$.

To measure the VDOS in Zr$_{46}$Cu$_{46}$Al$_8$ MG[54,55], INS measurements were performed at room temperature (RT) using ARCS[56] at the Spallation Neutron Source (SNS). The high flux brought forth by the source power, coupled with advances in neutron instrumentation, has enabled high-precision measurements to large momentum transfer, with fine energy resolution[57]. To facilitate the analysis of the INS data, the static structure factor $S(Q)$ was measured separately using the NOMAD[58] diffractometer at the SNS. Meanwhile, MD simulations were performed for the Zr$_{46}$Cu$_{46}$Al$_8$ MG with a realistic embedded atom method potential using the LAMMPS software package. The MD simulation results were benchmarked with the experimental $S(Q)$[55]. The vibration dynamics calculated by the MD simulations were analyzed in terms of the standard van Hove correlation function, which offered a direct comparison with INS measurements[55]. More details about the INS experiments and MD simulations can be found in Methods.

To investigate the momentum dependence of the BP, we first calculated the generalized $\boldsymbol{Q}$-dependent density of states (GDOS), $G(\boldsymbol{Q}, E)$, from the dynamic structure factor, $S(\boldsymbol{Q}, E)$, using the following equation[55,59,60]:

$$G(\boldsymbol{Q}, E) = e^{Q^2 u^2} \frac{E}{Q^2} \langle n \rangle S(\boldsymbol{Q}, E) \qquad (1)$$

where $\langle n \rangle = \left[1 - e^{-\frac{E}{k_B T}}\right]$ describes the Bose-Einstein statistics, $u^2$ is the average mean-square displacement, and $e^{-Q^2 u^2}$ describes the Debye-Waller factor, $k_B$ is the Boltzmann constant and $T$ the absolute temperature, respectively. Then, we divided $G(\boldsymbol{Q}, E)$ by $E^2$ to obtain $B(\boldsymbol{Q}, E) = G(\boldsymbol{Q}, E)/E^2$. Note that $B(\boldsymbol{Q}, E)$ can be compared to $g(E)/E^2$, which is commonly used for the analysis of the BP. In Debye model[38], the VDOS $g(E) \propto E^2$ and thus $g(E)/E^2$ becomes a constant at small $E$ values. As shown in Fig. 1a, the spectra of $B(Q, E)$ from INS measurements exhibit a high concentration of INS intensities at $E\sim5.5$ meV around $Q\sim2.8$ Å$^{-1}$. In MD simulations, Fig. 1b, the BP appears at $E\sim3.5$ meV, but the $Q$-dependence largely reproduces the experimental observations. Furthermore, in the vicinity of $E\sim5.5$ meV, the INS $B(Q, E)$ intensity exhibits a characteristic $Q$-dependence, which agrees well with the static structure factor $S(Q)$ measured separately with neutron diffraction, see Fig. 1c. This agreement is echoed by MD simulation results, Fig. 1d, which shows a nearly perfect correlation between $B(Q, E)$ and $S(Q)$ at the BP energy.

Next, we demonstrate that the excitation around 5.5 meV corresponds to the energy of the BP. As shown in Fig. 2a, the INS measured total GDOS, obtained by integrating over a $Q$ range of 2.1-8.5 Å$^{-1}$, was normalized by $E^2$. An excess $B(E)$ can be readily identified, centering around 5.5 meV for the $Zr_{46}Cu_{46}Al_8$ MG under study. To solidify the evidence, two other experimental techniques were employed to confirm the BP energy in $Zr_{46}Cu_{46}Al_8$ MG, i.e., Raman scattering and specific heat ($C_p$) measurements. As shown in Fig. 2b, a clear BP around 6.0 meV in the reduced Raman spectra can be identified. This is consistent with the INS measured BP at ~5.5 meV. Figure 2c plots the temperature dependence of $C_p$, in the form of $(C_p - \gamma T)/T^3$ vs $T$, where $\gamma$ is the Sommerfeld coefficient measuring the electronic contribution to $C_p$. As can be seen, $(C_p - \gamma T)/T^3$ does not stay constant as predicted by the Debye model for crystalline materials[61]. However, adding an Einstein oscillator term leads to a good fit to the $C_p$ data[61]. The fitted Einstein oscillator yields an average energy of ~6.5 meV, which is also consistent with the BP energies obtained above. With these three different measurements (i.e., INS, Raman scattering, and $C_p$), we can confirm that the excitation at ~5.5 meV in the INS spectra is indeed the BP.

Having established the BP, we performed a detailed analysis of the BP as a function of $Q$. For a given $Q$ value between 2.1 and 8.5 Å$^{-1}$, the constant $Q$ cut plots of $B(Q, E)$ all show a BP anomaly. Specifically, the BP peak position and peak height were extracted and examined, to explore the momentum dependence of the energy and intensity of the BP, respectively. Interestingly, the BP energy for different $Q$ values is quite similar, close to $E$ ~5.5 meV, which indicates that the BP is dispersionless, at least within the resolution of the ARCS instrument[56]. The lack of dispersion of the BP energy can also be seen in MD simulations at 3.5 meV, as shown in Fig. 1b. The BP energy of 3.5 meV in MD simulations is lower than the 5.5 meV determined by the INS experiment. This difference could be due to the extraordinarily high cooling rate used in MD simulations. Since the time scales in MD simulation are limited, fast cooling rate had to be used. In spite of the difference in the energy scale, the main features of the $B(Q, E)$ calculated by the MD simulations and INS experiment are largely the same, so a direct comparison can be made and the results are discussed together.

The dispersionless nature of the BP energy indicates that in glasses with disordered structures, the atoms could vibrate in resonance with the same energy scale. This observation goes on to suggest that the BP might be a characteristic excitation in amorphous alloys.

Although the BP energy at different $Q$ values is nearly constant at ~5.5 meV, the INS intensity at the BP energy varies strongly with $Q$. As shown in Fig. 1c, the momentum-dependent BP intensity shows a close one-to-one correspondence with $S(Q)$. This experimental observation was reproduced by MD simulations, as shown in Fig. 1d, demonstrating that the observed momentum dependence of the BP intensity is intrinsic. Apart from the results in Fig. 1c and 1d for Zr$_{46}$Cu$_{46}$Al$_8$ MG, which is an excellent glass former, similar results were also obtained for Zr$_{56}$Cu$_{36}$Al$_8$ MG with a marginal glass forming ability, in both INS measurements and MD simulations. Taken together, the data from INS measurements and MD simulations for two different alloys of different glass-forming abilities would indicate that the BP excitation is closely connected with the underlying amorphous structures.

To further investigate the nature of the BP excitation, constant *E* cut plots of the BP spectra at selected *E* values were extracted and examined. Figure 3a shows the constant *E* cut plots of *B*(*Q*, *E*) at *E*=15, 20, 25, and 40 meV (over Δ*E*=0.5 meV), respectively. The corresponding MD simulation results are shown in Fig. 3b. No distinct correlations can be seen between *S*(*Q*) and constant *E* cut of *B*(*Q*, *E*) at those energy values. For example, at *E*=25 meV, the maxima for the constant *E* cut of *B*(*Q*, *E*) and *S*(*Q*) appear at different *Q* positions. The consistency between the INS experiments and MD simulations (Fig. 1) further confirms that the BP is a unique excitation, different from phonons at other energies[55]. At the BP energy, the *B*(*Q*, *E*) scales with the static structure factor, *S*(*Q*).

A similar correlation of *B*(*Q*, *E*) with *S*(*Q*) at the BP energy was also observed in MD simulation for a 3-dimensional Lennard-Jones (3DLJ) binary glass. The simulation results confirm two main features of the BP: 1) the BP energy is largely dispersionless, and 2) the *B*(*Q*, *E*) intensity scales with *S*(*Q*). These findings, based on the simulation with a generic 3DLJ potential, suggest that the BP features reported here might be universal for glasses.

Based on INS and MD simulation results, an analytical form for the BP in amorphous materials can be formulated:

$$B_{BP}(\mathbf{Q}, E) = A * f(E - E_{BP})S(Q) \qquad (2)$$

where *A* is a scaling factor, and the function $f(E - E_{BP})$ is momentum-independent which has a peak at *E*=$E_{BP}$ that describes the dispersionless feature of the BP energy. It follows that the corresponding dynamic structure factor of the BP can be written as:

$$S_{BP}(\mathbf{Q}, E) = A * EQ^2 e^{-Q^2 u^2} \langle n \rangle^{-1} f(E - E_{BP}) S(Q) \tag{3}$$

Previous studies of the BP have primarily focused on the energy-dependent DOS, which is the Fourier transform of time-dependent velocity autocorrelation function, e.g., $g(\omega) = \frac{1}{2\pi} \int \frac{\psi_v(t)}{\psi_v(0)} exp(-i\omega t) dt$, where $\psi_v(t) = N^{-1} \sum_j \langle \mathbf{v}_{j,t} \cdot \mathbf{v}_{j,0} \rangle$. Guerdane and Teichler generalized the definition of the autocorrelation function by replacing the atomic velocity with fluctuations of atomic coordination number (CN), i.e., $\psi_{CN}(t) = N^{-1} \sum_j \langle (Z_{j,t} - \bar{Z}_j)(Z_{j,0} - \bar{Z}_j) \rangle$ [53]. The Fourier transform $\chi_{CN}(\omega) = \frac{1}{2\pi} \int \frac{\psi_{CN}(t)}{\psi_{CN}(0)} exp(-i\omega t) dt$ can be treated as a DOS, describing the oscillation frequency of atomic CN. They found that $\chi_{CN}(E)$ shows a clear peak at $E$ around $E_{BP}$, and thus concluded that the BP is related to time-dependent fluctuations of local environments of atoms.

To describe the $Q$-dependence, we consider the double Fourier transform of the spatial correlation of CN fluctuations between different atoms. The correlation function can be extended as:

$$\varphi_{CN}(\mathbf{Q}, t) = N^{-1} \sum_j \sum_k \langle (Z_{j,t} - \bar{Z}_j)(Z_{k,0} - \bar{Z}_k) e^{i\mathbf{Q} \cdot (r_{j,t} - r_{k,0})} \rangle \tag{4}$$

Then, the generalized momentum-dependent DOS for CN fluctuation follows:

$$\Omega_{CN}(\mathbf{Q}, \omega) = \frac{1}{2\pi} \int \frac{\varphi_{CN}(\mathbf{Q}, t)}{\varphi_{CN}(\mathbf{Q}, 0)} e^{-i\omega t} dt \tag{5}$$

As shown in Fig. 4a, $\chi_{CN}(E)$ exhibits a clear peak around the BP energy in our $Zr_{46}Cu_{46}Al_8$ MG, which is consistent with a previous study[53]. However, as Fig. 4b shows, the momentum-dependent $\Omega_{CN}(\mathbf{Q}, E)$ did not produce the $S(Q)$-like feature observed in INS experiment and MD simulations (Fig.1).

To further explore the underlying physics of the BP, we went on to study the spatial and temporal fluctuations of several other parameters, such as the nearest-neighbor (NN) distance and local atomic strain, by using Eq.(4)&(5). Previous studies have shown that the fluctuation of the NN distance is closely linked to the variation of force constants which has been assumed to be the origin of BP[21]. However, as Fig. 4c and 4a demonstrate, it captures neither the BP energy nor the $S(Q)$-like feature of BP intensity. The local atomic strain can be decomposed into three terms: the non-affine strain, the volumetric strain, and the shear strain. Their double Fourier transforms are shown in Figs. 4(d-f). Interestingly, only the shear strain fluctuation produces both the energy and momentum dependence of the BP as observed by our INS experiment and MD simulations (Fig. 1). Fluctuations of the non-affine strain and volumetric strain did not show such correlations. These simulation results strongly suggest that BP is intimately associated with the fluctuation of the local shear strain around each atom due to topological disorder. In the future, fluctuations of other order parameters could also be checked in a similar way. The above analysis also serves as a reminder that an apparent peak in the energy-dependent DOS does not necessarily correspond to the BP.

We noticed that earlier published works adopted different expressions of $S(Q, E)$, namely $S(Q, E)*E/Q$, or $S(Q, E)*<n>/(E*Q)$ for BP [17,33,60], to examine the $Q$-dependence of the INS data. We also calculated these quantities and compared the results with $B(Q, E)$. While the explicit values are different, a $Q$-dependent feature can be seen for $S(Q, E)$ and $S(Q,E)*<n>/(E*Q)$ around the BP energy.

In summary, by interrogating the momentum dependence of INS and MD simulation results, we have established important characteristics of the BP in amorphous alloys. We found that the BP excitation is essentially dispersionless, i.e., with a fixed energy. In addition, a one-to-one correspondence was found between the INS intensity (represented in the form of $B(Q, E)$) at the BP excitation energy and the static structure factor $S(Q)$, and this was reproduced by MD simulations with calibrated empirical potentials. These features could be fingerprints to distinguish the origin of the BP in amorphous materials from those observed in other systems. A simple analytical expression describing the dynamic structure factor of the BP excitation has been formulated for the MG under study. Inspired by the experimental observations, we found, through MD simulations, that fluctuations of the local atomic shear strain reproduce both the dispersionless feature of the BP energy and the momentum-dependent BP intensity. We also ruled out four other models involving fluctuations by the local CN, the NN distance, the non-affine strain, and the volumetric strain. Although the present study was demonstrated for MGs, the same methodology could be applied to other types of glasses. The powerful combination of INS and MD not only shed light on the nature of the BP but also pointed out the importance of considering shear strain fluctuations for understanding the dynamic properties of amorphous materials.

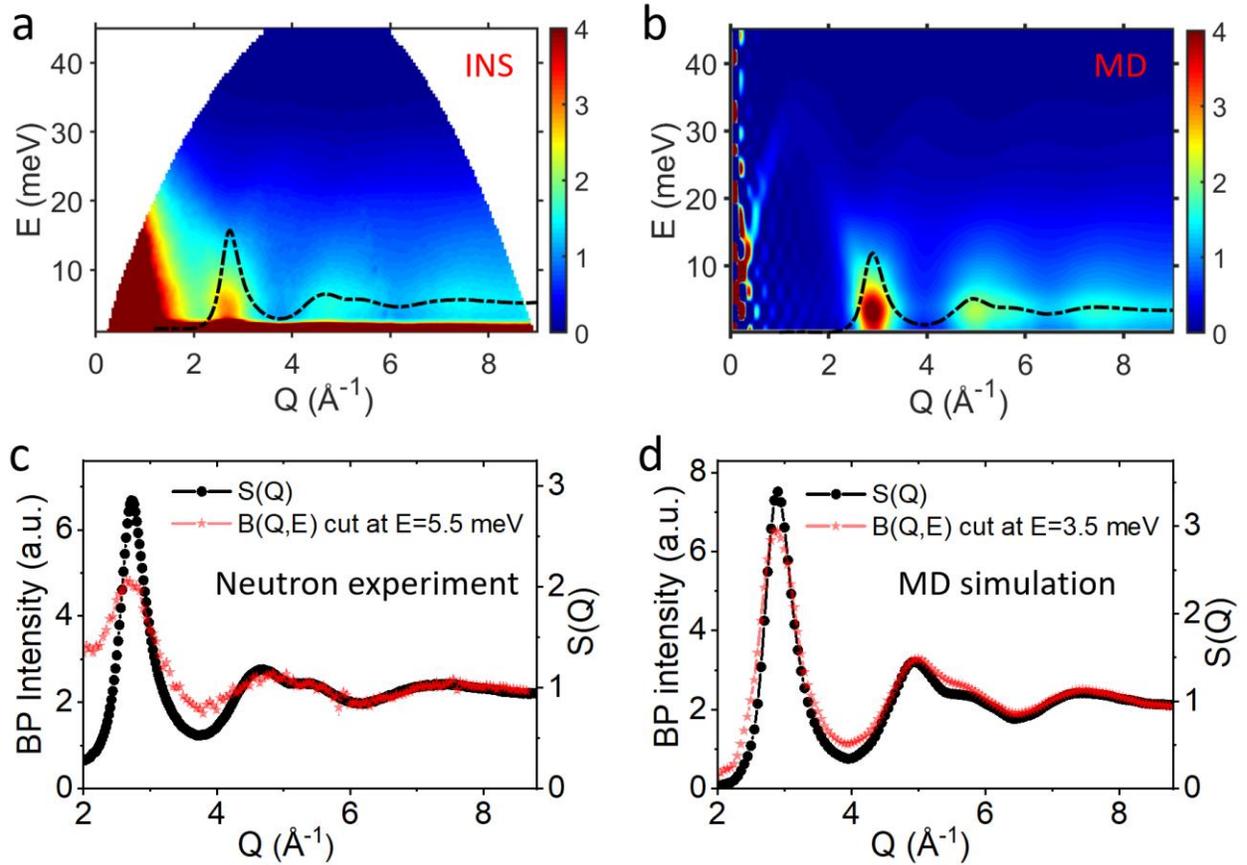

**Fig. 1 | Momentum dependence of the BP in $Zr_{46}Cu_{46}Al_8$ MG.** (**a**) Generalized $Q$-dependent density of states normalized by $E^2$, $B(Q, E)$, measured by INS (see text). The strong INS intensity close to $E=0$ was due to the finite instrument resolution and the elastic scattering. (**b**) MD simulation results for $B(Q, E)$. The INS data show a well-defined BP around 5.5 meV which is largely dispersionless as a function of $Q$. The static structure factor $S(Q)$ is superimposed (the black dash lines in a and b). (**c**) and (**d**) The constant $E$ cut plots at the BP energy (over $\Delta E=0.5$ meV) were superimposed with the experimental $S(Q)$. The INS intensity at the BP energy exhibits an excellent correlation with $S(Q)$.

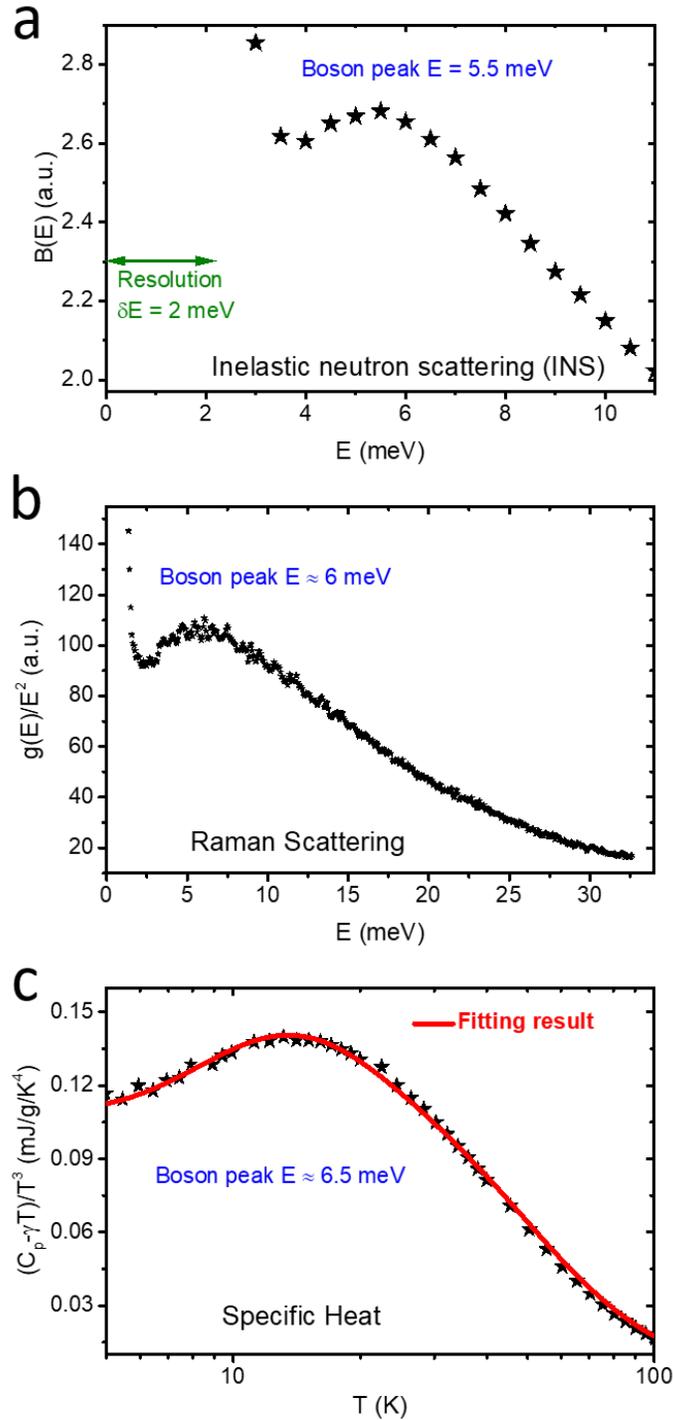

**Fig. 2 | BP in $Zr_{46}Cu_{46}Al_8$ MG.** (**a**) The INS measured vibrational density of states normalized by $E^2$, $B(E)$, showing a BP at ~5.5 meV. Within the Debye model, $B(E)$ would appear as a straight line at small $E$ values. (**b**) The vibrational density of states normalized by $E^2$, $g(E)/E^2$, obtained from the Raman scattering data, showing a BP around 6.0 meV. (**c**) The specific heat data presented as $(C_p - \gamma T)/T^3$ vs $T$, which was fitted by the Einstein oscillator model. The fitted energy value for the Einstein oscillator mode was around 6.5 meV.

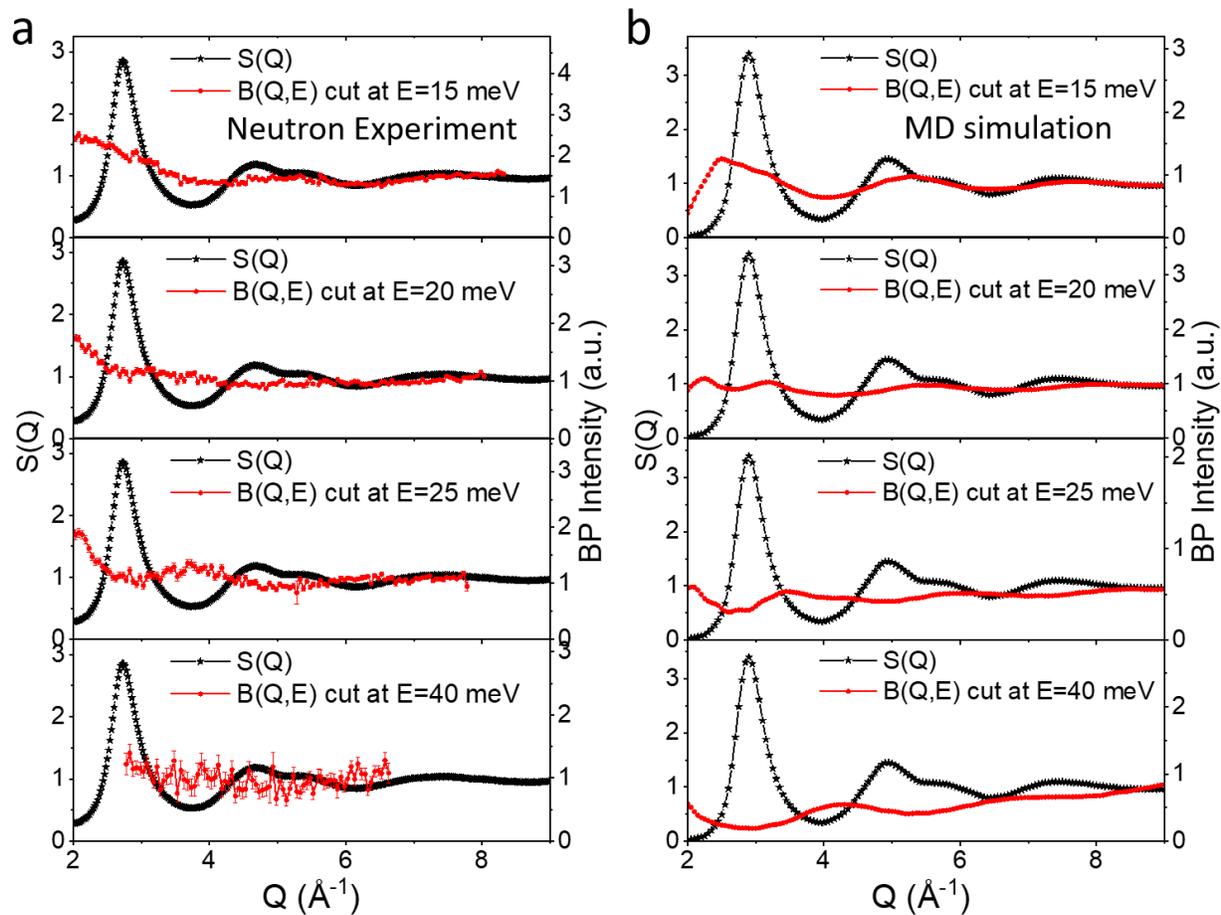

**Fig. 3 | Constant *E* cut plots of *B*(*Q*, *E*) along with the static structure factor *S*(*Q*) in Zr$_{46}$Cu$_{46}$Al$_8$ MG.** (**a**) INS measurements. (**b**) MD simulations. Unlike at the BP excitation energy, the *B*(*Q*, *E*) intensity at the indicated *E* values are not correlated with *S*(*Q*).

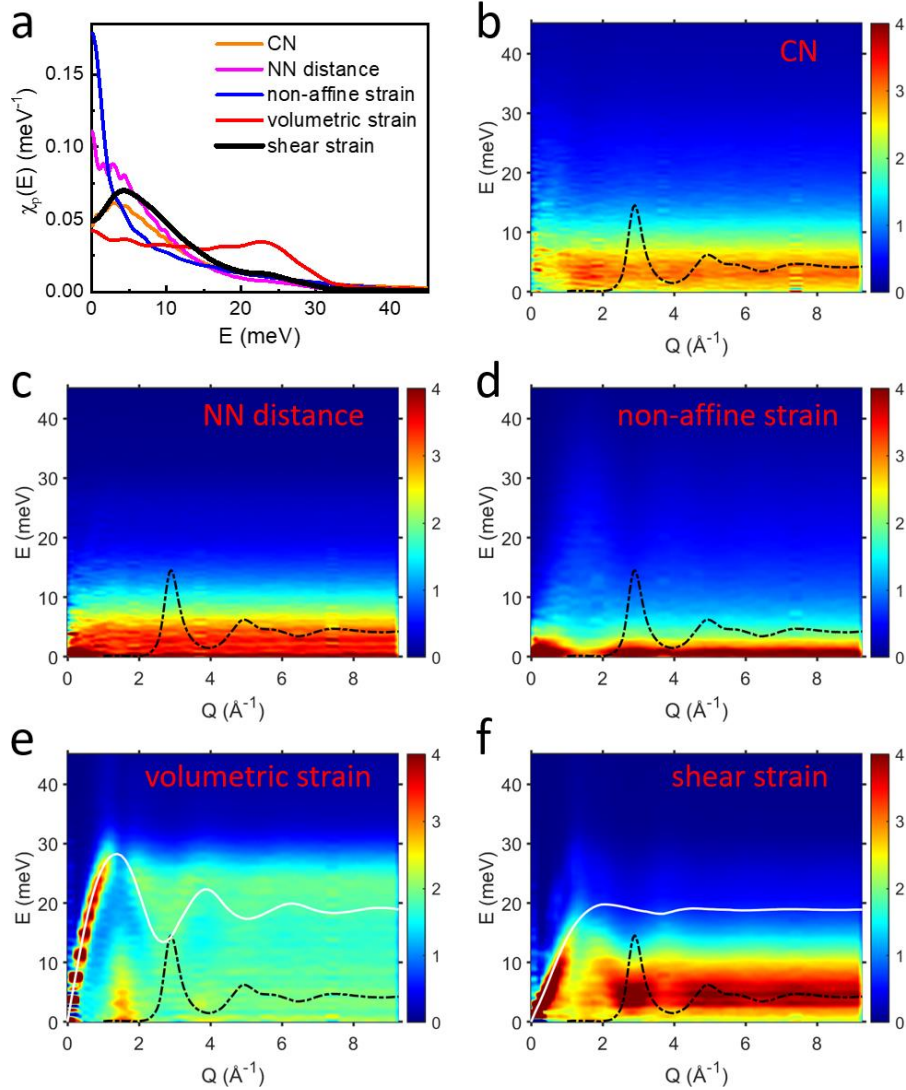

**Fig. 4 | The $Q$-dependent frequency spectrum resulting from various structure fluctuations in Zr$_{46}$Cu$_{46}$Al$_8$ MG.** (**a**) The $Q$-integrated frequency spectrum which reflects the density of states. (**b**) The local coordination number (CN) fluctuation model, which shows a dispersionless BP but not $S(Q)$-like BP intensity. (**c**) The nearest-neighbor (NN) distance fluctuation model, which shows neither a dispersionless BP nor $S(Q)$-like BP intensity. (**d**) The non-affine strain model, which shows $S(Q)$-like BP intensity but not the gapped dispersionless BP energy. (**e**) The volumetric strain model, which shows none of the observed BP features. (**f**) The shear strain model, which shows a dispersionless BP and $S(Q)$-like BP intensity at $E$~4.3 meV, capturing the main features observed in the INS experiment and MD simulations as shown in Fig.1. Note that the volumetric strain and shear strain model respectively produced the longitudinal and transverse phonon dispersions (white curves in (**e**) and (**f**)) which we reported in Ref[55]. The $S(Q)$ was superimposed as black dash lines.